\documentclass[aps,twocolumn,aps,floatfix]{revtex4}
\usepackage{graphicx}
\usepackage{bm}
\usepackage{hyperref}
\def\he#1{$^{#1}$He}
\def\hm2{\frac{\hbar^2}{2m}}
\def\vecr{ {\bf r}}

\begin{document}
\title{Layer-- and bulk roton excitations of $^4$He in porous media}
\author{V. Apaja and E. Krotscheck}
\affiliation{Institut f\"ur Theoretische Physik, Johannes Kepler
Universit\"at, A 4040 Linz, Austria}

\begin{abstract}

We examine the energetics of bulk and layer--roton excitations of \he4
in various porous medial such as aerogel, Geltech, or Vycor, in order
to find out what conclusions can be drawn from experiments on the
energetics about the physisorption mechanism. The energy of the
layer--roton minimum depends sensitively on the substrate strength,
thus providing a mechanism for a direct measurement of this
quantity. On the other hand, bulk-like roton excitations are largely
independent of the interaction between the medium and the helium
atoms, but the dependence of their energy on the degree of filling
reflects the internal structure of the matrix and can reveal features
of \he4 at negative pressures. While bulk-like rotons are very similar
to their true bulk counterparts, the layer modes are not in close
relation to two--dimensional rotons and should be regarded as a third,
completely independent kind of excitation.

\end{abstract}
\maketitle

\section{Introduction}

Collective excitations of superfluid helium confined in various porous
media have been studied by neutron scattering since early 90's, and by
now a wealth of information about helium in aerogel, Vycor and Geltech
has been collected \nocite{sokol-gibbs-stirling-azuah-adams-96,
dimeo-etal-97,dimeo-sokol-anderson-stirling-adams-98,
gibbs-sokol-stirling-azuah-adams-97,GFP98,
FPGM00,GPFCDS00,PFGMBCS01,PGFBAMS02,LauterAerogel}
\cite{sokol-gibbs-stirling-azuah-adams-96}-\cite{LauterAerogel}. Aerogel
is an open gel structure formed by silica strands (SiO$_2$). Typical
pore sizes range from few \AA\ to few hundred \AA\ , without any
characteristic pore size. Vycor is a porous glass, where pores form
channels of about 70~\AA\ diameter. Geltech resembles aerogel, except
that the nominal pore size is 25~\AA\ \cite{PGFBAMS02}.

Liquid $^4$He is adsorbed in these matrices in the form of atomic
layers, the first layer is expected to be solid; on a more strongly
binding substrate, such as graphite, one expects two solid
layers. Energies and lifetimes of phonon--roton excitations for
confined $^4$He are nearly equal to their bulk superfluid $^4$He
values \cite{anderson-etal-99}, but differences appear at partial
fillings. The appearance of ripplons is tied to the existence of a
free liquid surface; neutron scattering experiments show clearly their
presence in adsorbed films \cite{LauterJLTP,LauterPRL,LauterAerogel}
with few layers of helium.

An exclusive feature of adsorbed films is the appearance of ``layer
modes''. The existence of such excitations has been proposed in the
seventies \cite{Padmore2Droton,GoetzeLuecke76} from theoretical
calculations of the excitations of two--dimensional \he4 and
comparison with specific heat data. Direct experimental evidence for
the existence of collective excitations below the roton minimum has
first been presented by Lauter and collaborators
\cite{LauterPRL,filmexpt}, identification of these excitations with
longitudinally polarized phonons that propagate in the liquid layer
adjacent to the substrate has been provided by microscopic
calculations of the excitations of films
\cite{filmexc,ApajaKroSqueezed}.

\begin{figure}
\includegraphics[width=0.48\textwidth]{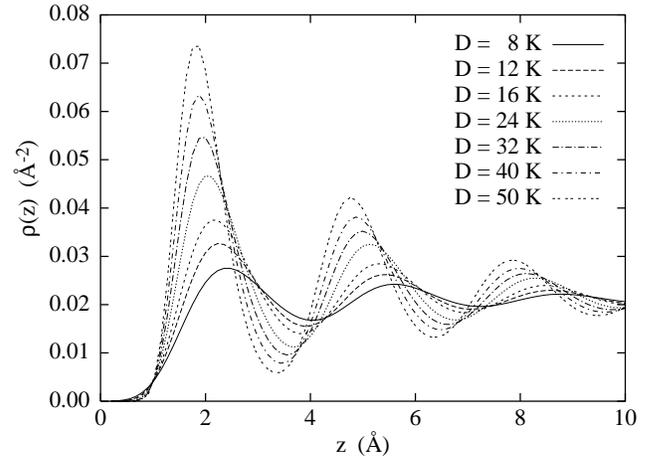}
\caption{The density profiles of the first three layers are
shown as a function of the depth $D$ of the substrate potential.
The substrate is at $z<0$.
}
\label{fig:profiles}
\end{figure}

In an experimental situation, the topology gives rise to non--uniform
filling of the pores. But from the theoretical point of view different
materials are characterized solely by their substrate potentials,
because as long as the wavelength of the excitation in concern is much
shorter than any porosity length-scale, the topology of the confining
matrix is immaterial. We therefore examine the energetics of the
layer--roton as a function of the substrate--potential strength which
determines, in turn, the areal density in the first liquid layer.  For
that purpose, we have carried out a number of calculations of the
structure of helium films as a function of potential strength.  The
microscopic theory behind these calculations is described in
Ref. \cite{filmstruc}.  Our model assumes the usual 3-9 potential
\begin{equation}
U_3(z) = \left[{4 C_3^3\over 27 D^2}\right]{1\over z^9} - {C_3\over z^3}\,;
\label{eq:Usub}
\end{equation}
we have varied the potential strength $D$ from 8~K to 50~K and the
range $C_3$ from 1000~K~\AA$^3$ to 2500~K\AA$^3$. In all cases, we have
considered rather thick films of an areal density of 0.45~\AA$^{-2}$.
Fig. \ref{fig:profiles} shows density profiles for these potential
strengths close to the substrate; the density profiles are practically
independent of the potential range $C_3$.

\section{Theory of excitations}
\label{sec:theory}

To introduce excitations to the system one applies a small,
time--dependent perturbation that momentarily drives the quantum
liquid out of its ground state. Generalizing the Feynman--Cohen
wave function \cite{FeynmanBackflow}, we write the
excited state in the form
\begin{equation}
\left|\Psi(t)\right\rangle = {e^{-iE_0 t/\hbar} \,
        e^{{1\over 2}\delta U(t)}\left|\Psi_0\right\rangle\over
        \left[\left\langle\Psi_0\left|e^{\,\Re e\delta U(t)}
        \right|\Psi_0\right\rangle\right]^{1/2}}\,,
\end{equation}
where $\left|\Psi_0\right\rangle$ is the exact or an optimized
variational ground state, and the excitation operator is
\begin{equation}
\delta U(t) = \sum_i\delta u_1({\bf
r}_i;t) + \sum_{i<j} \delta u_2(\vecr_i,\vecr_j;t)+\ldots\,.
\label{eq:deltaU}
\end{equation}
The time--dependent excitation functions $\delta
u_n(\vecr_1,\ldots,\vecr_n;t)$ are determined by an
action principle
\begin{equation}
        \delta \int ^{t_1}_{t_0} dt
        \left\langle \Psi (t) \left\vert H-i\hbar {\partial \over
\partial t} + U_{\rm ext}(t)\right\vert \Psi (t) \right\rangle   = 0\,,
\label{eq:S}
\end{equation}
where $U_{\rm ext}(t)$ is the weak external potential driving the
excitations.  The truncation of the sequence of fluctuating
correlations $\delta u_n$ in Eq.~(\ref{eq:deltaU}) defines the level
of approximation in which we treat the excitations. One recovers the
Feynman theory of excitations \cite{Feynman} for non--uniform systems
\cite{ChangCohen} by setting $\delta u_2(\vecr_1,\ldots \vecr_n;t) =
0$ for $n\ge 2$. The two--body term $\delta u_2(\vecr_1,\vecr_2;t)$
describes the time--dependence of the short--ranged correlations. It
is plausible that this term is relevant when the wavelength of an
excitation becomes comparable to the interparticle
distance. Consequently, the excitation spectrum can be quite well
understood \cite{JacksonSkw,Chuckphonon,VesaMikkou2} by retaining only
the time--dependent one-- and two--body terms in the excitation
operator (\ref{eq:deltaU}). The simplest non--trivial implementation
of the theory leads to a density--density response function of the
form \cite{filmexc}
\begin{eqnarray}
        &&\chi({\bf r},{\bf r}',\omega) = \\
	&&\sqrt{\rho({\bf r})}
        \sum_{st}\phi^{(s)}({\bf r})
        \left[G_{st}(\omega)+
                G_{st}(-\omega)\right]
        \phi^{(t)}({\bf r}')\sqrt{\rho({\bf r}')}\nonumber
\label{eq:CBFresponse}
\end{eqnarray}
where the $\phi^{(s)}({\bf r})$ are Feynman excitation functions, and
\begin{equation}
G_{st}(\omega) =
\left[\hbar[\omega - \omega_s+i\epsilon]\delta_{st}
+ \Sigma_{st}(\omega)\right]^{-1}
\label{eq:Gdef}
\end{equation}
the phonon propagator. The fluctuating pair correlations
give rise to the dynamic self energy correction \cite{filmexc},
\begin{equation}
\Sigma_{st}(\omega) =
\frac{1}{2}\sum_{mn}\frac{V_{mn}^{(s)}V_{mn}^{(t)}}
{\hbar(\omega_m + \omega_n - \omega)}\,.
\label{eq:selfen}
\end{equation}
Here, the summation is over the Feynman states $m, n$; they form a
partly discrete, partly continuous set due to the inhomogeneity of the
liquid.  The expression for the three--phonon coupling amplitudes
$V_{mn}^{(s)}$ can be found in Ref.~\onlinecite{filmexc}. This self
energy renormalizes the Feynman ``phonon'' energies $\hbar\omega_n$, and
adds a finite lifetime to states that can decay.  The form of the self
energy given in Eq.~(\ref{eq:selfen}) is the generalization of the
correlated basis functions (CBF)~\cite{JacksonSkw,Chuckphonon} theory
to inhomogeneous systems.  As a final refinement of the theory, we
scale the Feynman energies $\omega_n$ appearing in the energy
denominator of the self energy given in Eq.~(\ref{eq:selfen}) such
that the roton minimum of the spectrum used in the energy denominator
of Eq. (\ref{eq:selfen}) agrees roughly with the roton minimum
predicted by the calculated $S({\bf k}, \omega)$. This is a
computationally simple way of adding the self energy correction to the
excitation energies in the denominator of Eq. (\ref{eq:selfen}). We
shall use this approximation for the numerical parts of this paper.

\section{Layer excitations}

\begin{figure}
\includegraphics[width=0.48\textwidth]{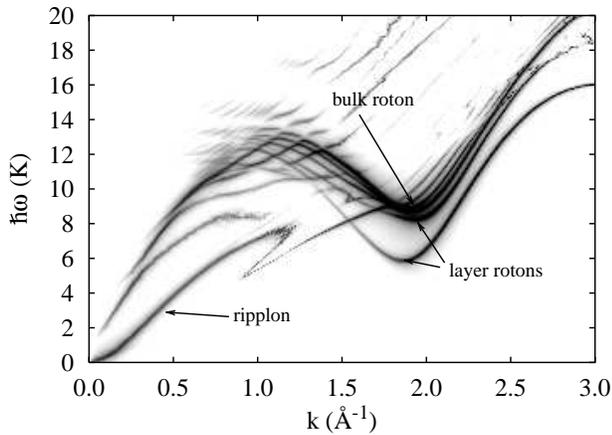}
\caption{The figure shows the map of the dynamic structure
function $S({\bf k},\omega)$ for a \he4 film for the potential strength $D =
32\,$K. The two layer--rotons, the bulk roton, and the ripplon are
indicated by arrows.}
\label{fig:skwplot}
\end{figure}

Layer phonons are identified by a transition density that is localized
in the first liquid layer of the system. They appear in the dynamic
structure function $S({\bf k},\omega)$ as a peak below the roton
minimum.  A grayscale map of a typical dynamic structure function is
shown in Fig. \ref{fig:skwplot}, we have for clarity chosen a momentum
transfer parallel to the substrate; neutron scattering at other angles
would broaden the roton minima \cite{ApajaKroSqueezed}.  The figure
shows in fact one bulk and two layer--roton minima, but the higher
one, which corresponds to an excitation propagating in the second
liquid layer, has an energy too close to the bulk roton to be
experimentally distinguishable.

\begin{figure}
\vbox{
\includegraphics[width=0.4\textwidth]{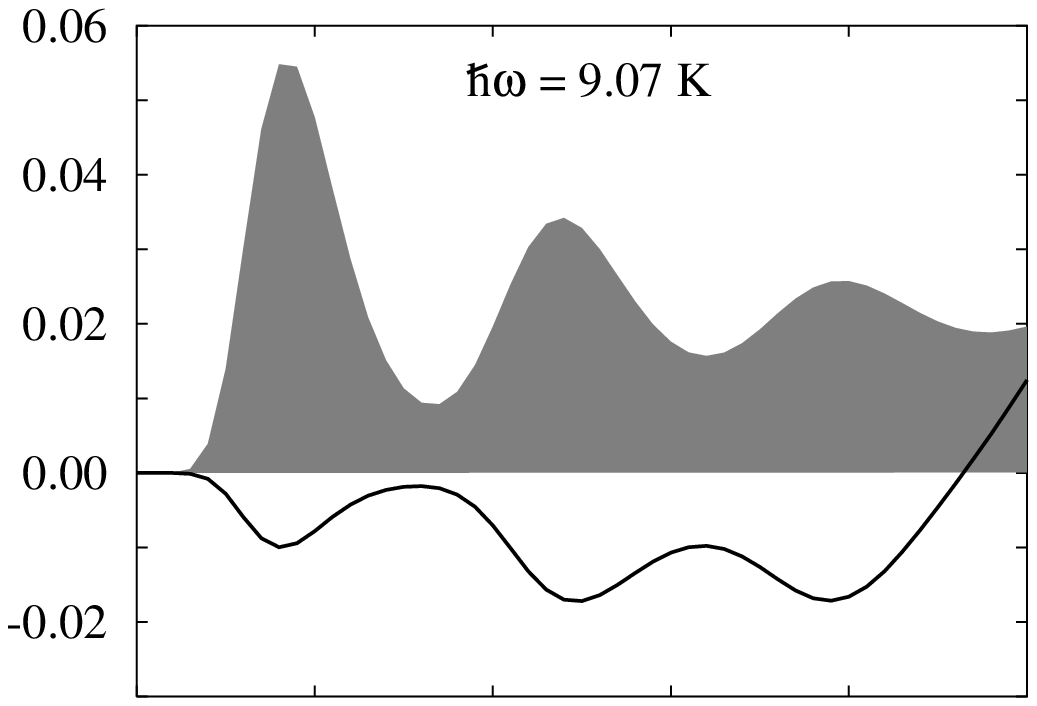}\\
\includegraphics[width=0.4\textwidth]{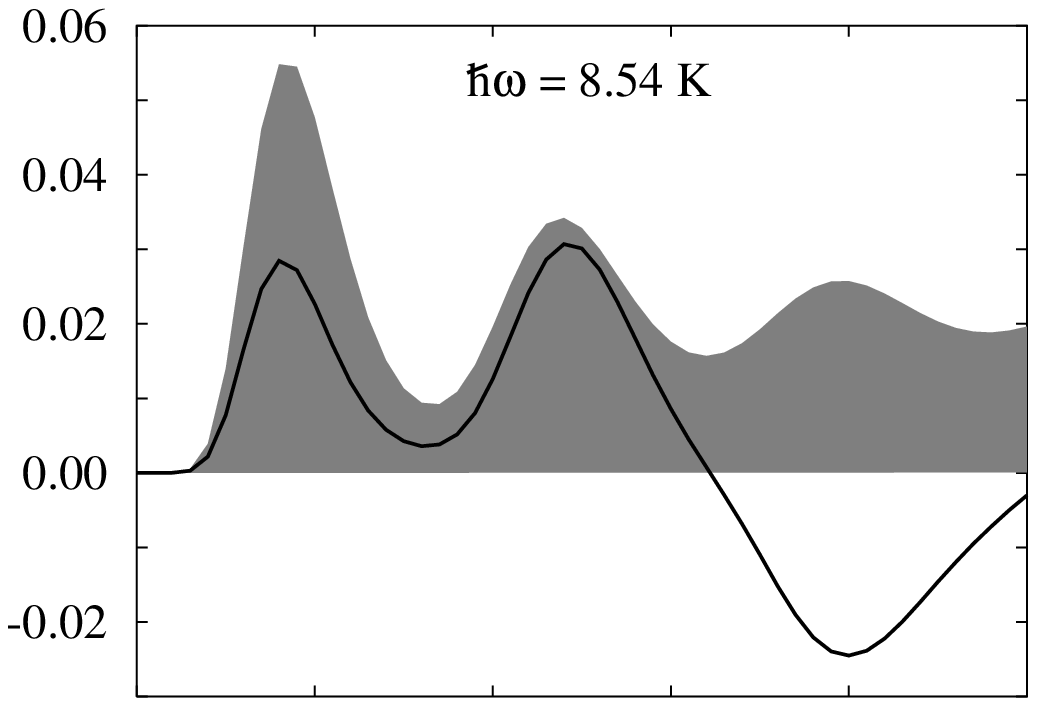}\\
\includegraphics[width=0.4\textwidth]{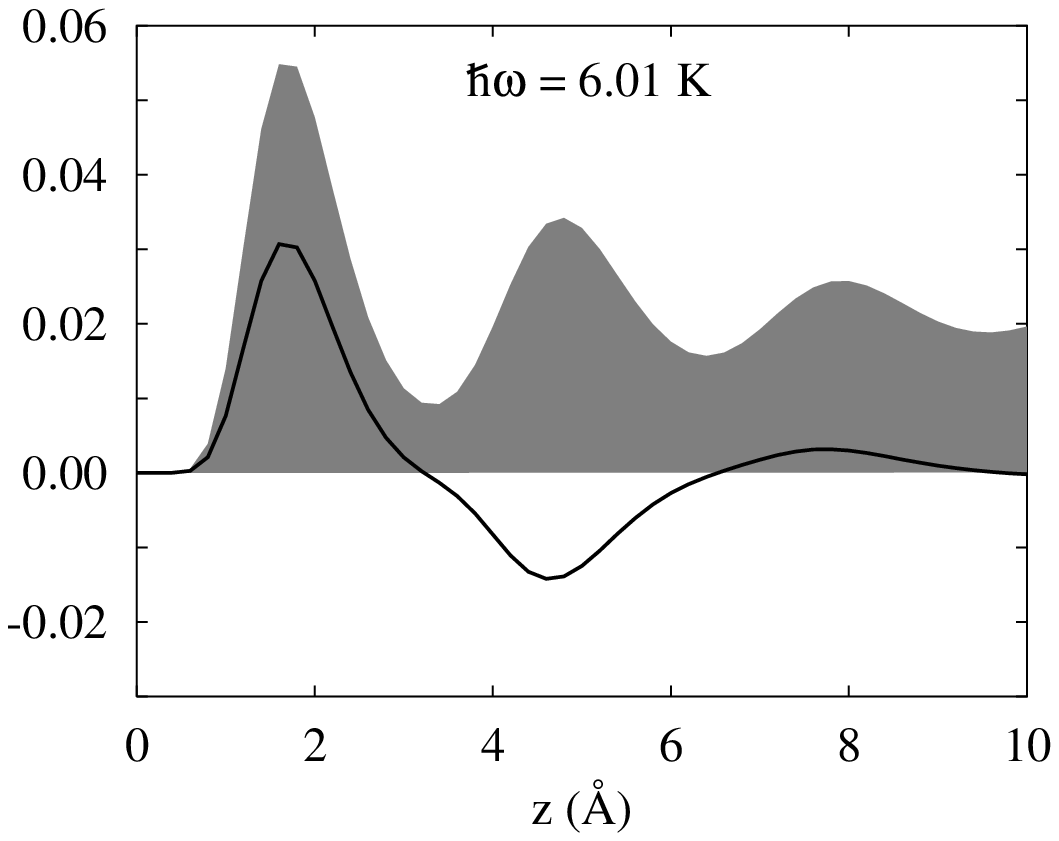}
}
\caption{The transition densities of the three lowest excitation are
shown for $D=32$~K and $C_3=1500$~K\AA$^3$, normalized to the same
maximum value. For comparison, the density profile of the film is
shown as a gray--shaded area.}
\label{fig:trans}
\end{figure}

The transition densities corresponding to the three pronounced
excitations at $k = 1.8\,$\AA$^{-1}$ are depicted in
Fig.~\ref{fig:trans}. Clearly, the two ``layer--modes'' are actually
located in the two first layers adjacent to the substrate whereas the
``bulk'' mode is spread throughout the system. However, the figure
also shows that the notion that the wave propagates in the first or the
second layer is also not quite accurate: The lowest mode also has some
overlap with the second layer, but especially the second mode spreads
over both layers.

\begin{figure}
\includegraphics[width=0.48\textwidth]{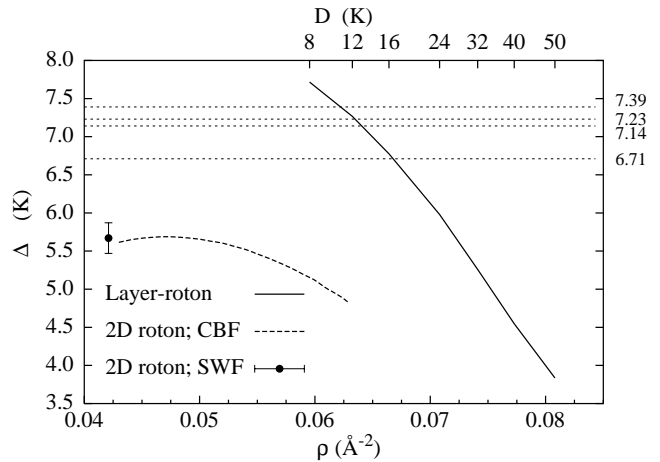}
\caption{The figure shows the energy of the roton in two dimensions,
(long--dashed line) and the energy of the layer roton in a film (solid
line) as a function of areal density. The areal densities of the
layers were obtained by integrating the density up to the first
minimum, the upper horizontal axis shows the corresponding values of
the well-depth $D$. Also shown is the energy of a two--dimensional
roton obtained with shadow wave functions \cite{Reatto2DRoton} at the
density of 0.0421\AA$^{-2}$. The short--dashed horizontal lines show
experimental values of the roton energy on aerogel
\cite{FPGM00,LauterAerogel} (7.39 and 7.14~K),
Vycor\cite{GPFCDS00,PFGMBCS01} (7.23 K), and Geltech\cite{PGFBAMS02}
(6.71 K), their energies are marked on the right margin.}
\label{fig:rotonenergy}
\end{figure}

We have carried out two independent calculations of the
two--dimensional roton excitation: First, we calculated the roton
energy as a function of the density for a rigorously two dimensional
liquid. We can assess the accuracy of our predictions with the
shadow--wave--function calculation of $S({\bf k},\omega)$ of
Ref. \onlinecite{Reatto2DRoton}, who obtained a roton energy of $5.67
\pm 0.2\,$K at the equilibrium density of $n = 0.0421\,$\AA$^{-2}$.
Second, we have calculated the dynamic structure function $S({\bf
k},\omega)$ in the relevant momentum region for the above family of
substrate potentials. The results are compiled in
Fig. \ref{fig:rotonenergy} where we also collect several experimental
values.

Although exactly the same method has been used for the computation of
the purely two-dimensional system and for the films, the results are
quite different. We have obtained for the film calculation an
effective layer density by integrating the three-dimensional densities
shown in Fig. \ref{fig:profiles} to the first minimum. This is
evidently not very well defined for the weakly bound systems, but it
is not legitimate either for the case of strong binding where the
first layer is well defined. In fact, the integrated density for the
strongest substrate is 0.08~\AA$^{-2}$, which is well beyond the
solidification density of the purely two--dimensional system.
Evidently, the zero--point motion in $z$ direction can effectively
suppress the phase transition. We make therefore three conclusions:
(i) The position of the layer roton minimum is indeed a sensitive
measure for the strength of the substrate potential, (ii) purely
two--dimensional models are manifestly inadequate for their
understanding, and, hence, (iii) purely two--dimensional models are
also questionable for interpreting thermodynamic data of adsorbed
films.

\section{Bulk excitations}

With one exception \cite{FPGM00}, the bulk roton energy in porous
media have been reported to be practically identical to that in the
bulk liquid, Ref. \onlinecite{FPGM00} reports a slight increase of the
roton energy in aerogel at partial filling.  A roton energy above the
one of the bulk liquid can be explained by assuming that the density
of the helium liquid in the medium is below that of the bulk
liquid. This can, in turn, be qualitatively explained by the cost in
energy to form a surface.

\begin{figure}
\includegraphics[width=0.48\textwidth]{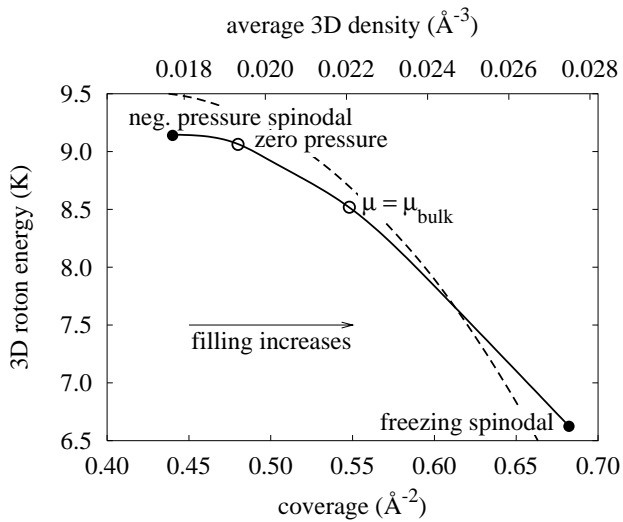}
\caption{The figure shows the energy of the bulk roton in a gap of 25
\AA\ width as a function of the two--dimensional coverage and the
corresponding average density. The points of zero pressure, the
low--density and the high--density spinodal points, and the bulk
chemical potential are indicated as points. The dashed line shows the
roton energy of a purely 3D calculation.}
\label{fig:3D-roton}
\end{figure}

To be quantitative, we have performed calculations of the energetics
and structure of \he4 in a gap between attractive silica walls
\cite{ApajaKroSqueezed} and obtained the energy of the bulk roton
(c.f. Fig. \ref{fig:skwplot}) as a function of filling.
Fig. \ref{fig:3D-roton} shows, as a typical example, the roton
energetics in in a gap of 25~\AA\ width. The independent parameter is
the areal density $n$, the corresponding three--dimensional density
was obtained by averaging the density profile over the full volume. It
is seen that the equilibrium density is well below the bulk value. In
other words, the roton energy in a confined liquid should correspond
to the one of a liquid that would, without confinement, have a
negative pressure.  The energy increase of the roton minimum found in
this model is about 0.5 K, which is consistent with the experiments of
Ref. \onlinecite{FPGM00}.

To verify this interpretation of the data, it would be very useful to
have comparable measurements for porous media with a more uniform
distribution of pore sizes. In particular, comparably small pores
should allow to densities that are even below the bulk spinodal
density \cite{ApajaKroHectorite}, thus facilitating experiments on
\he4 in density areas that were up to now inaccessible.

\begin{acknowledgments}
This work was supported by the Austrian Science Fund (FWF) under
project P12832-TPH. 
\end{acknowledgments}


\end{document}